\documentclass[11pt,twoside]{article}
\usepackage{asp2010}

\usepackage{color}
\newcommand{\about}{$\sim\!\!$~}
\newcommand{\kms}{km s$^{-1}$}
\def\arcdeg{\hbox{$^\circ$}}
\def\dm15{\ifmmode{\Delta m_{15}}\else{$\Delta m_{15}$}\fi}

\def\reff@jnl#1{{\rm#1\/}}
\def\apj{\reff@jnl{ApJ}}
\def\aj{\reff@jnl{AJ}}                  % Astronomical Journal
\def\araa{\reff@jnl{ARA\&A}}            % Annual Review of Astron and Astrophys
\def\apj{\reff@jnl{ApJ}}                % Astrophysical Journal
\def\apjl{\reff@jnl{ApJ}}               % Astrophysical Journal, Letters
\def\apjs{\reff@jnl{ApJS}}              % Astrophysical Journal, Supplement
\def\aap{\reff@jnl{A\&A}}               % Astronomy and Astrophysics
\def\mnras{\reff@jnl{MNRAS}}            % Monthly Notices of the RAS
\def\prd{\reff@jnl{Phys.Rev.D}}         % Physical Review D
\def\prl{\reff@jnl{Phys.Rev.Lett}}      % Physical Review Letters
\def\pasp{\reff@jnl{PASP}}              % Publications of the ASP
\def\nat{\reff@jnl{Nature}}             % Nature
\def\iaucirc{\reff@jnl{IAU~Circ.}}% 
\def\aaps{\reff@jnl{A\&AS}}% 
\def\pasa{\reff@jnl{PASA}}% 
\newcommand\actaa{\ref@jnl{Acta Astron.}}%
\aspvolume{491} % Insert volume number from ASPCS
\aspcpryear{2014} % Insert year received from ASPCS
\aspvoltitle{Fifty Years of Wide Field Studies in the Southern Hemisphere} 
\aspvolauthor{S. Points and A. Kunder, eds.} 

\aspLCCN{2014954058}
\aspISBN{978-1-58381-866-4}
\aspeISBN{978-1-58381-867-1}

\asphalftitle{FIFTY YEARS OF WIDE FIELD STUDIES IN THE SOUTHERN HEMISPHERE:
RESOLVED STELLAR POPULATIONS OF THE GALACTIC BULGE AND MAGELLANIC CLOUDS}

\aspcoverinfo{The Cerro Tololo Inter-American Observatory (CTIO) 4m Blanco
telescope dome below the central part of the Milky Way; the Magellanic Clouds 
are just above the horizon in the right corner. Image courtesy of T.
Abbott/NOAO/AURA/NSF.}

\aspmaintitle{FIFTY YEARS OF WIDE FIELD STUDIES IN THE SOUTHERN HEMISPHERE:
RESOLVED STELLAR POPULATIONS OF THE GALACTIC BULGE AND MAGELLANIC CLOUDS} 

\aspmtgtype{conference celebrating CTIO's 50th anniversary} 
\aspmtgvenue{La Serena, Chile} 
\aspmtgwhen{06--09 May 2013} 

\aspfirsteditor{S.~Points}
\aspfirstaffil{Cerro Tololo Inter-American Observatory (CTIO), Casilla 603, 
La Serena, Chile}

\aspsecondeditor{A.~Kunder}
\aspsecondaffil{Cerro Tololo Inter-American Observatory (CTIO), Casilla 603,
La Serena, Chile\\
Leibniz-Institut f\"{u}r Astrophysics (AIP), An der Sternwarte 16, 14482 
Potsdam, Germany}

\begin{document}

\title{Light Echoes of Ancient Transients with the Blanco CTIO 4m Telescope}
\author{Armin Rest$^1$, B. Sinnott$^2$, D. L. Welch$^2$, J. L. Prieto$^3$, F.~B. Bianco$^4$, T. Matheson$^5$, R.~C. Smith$^{5,6}$ and N.~B. Suntzeff$^{7,8}$
\affil{$^1$STScI, 3700 San Martin Dr., Baltimore, MD 21218, USA}
\affil{$^2$Department of Physics and Astronomy, McMaster University, Hamilton, Ontario L8S 4M1, Canada}
\affil{$^3$Department of Astrophysical Sciences, Princeton University, 4 Ivy Lane, Princeton, NJ 08544, USA}
\affil{$^4$Center for Cosmology and Particle Physics, New York University, 4 Washington Place, New York, NY 10003, USA}
\affil{$^5$National Optical Astronomy Observatory, 950 N. Cherry Avenue, Tucson, AZ 85719, USA}
\affil{$^6$Cerro Tololo Inter-American Observatory, Colina el Pino S/N, La Serena, Chile}
\affil{$^7$Dept of Physics and Astronomy, Texas A\&M University, College Station, TX 77843, USA}
\affil{$^8$Mitchell Institute for Fundamental Physics and Astronomy, Texas A\&M University, College Station, TX 77843, USA}
}

\begin{abstract}
For over a century, light echoes have been observed around variable
stars and transients. The discovery of centuries-old light echoes from
supernovae in the Large Magellanic Cloud has allowed the spectroscopic
characterization of these events using modern instrumentation, even in
the complete absence of any visual record of those events.  Here we
review the pivotal role the Blanco 4m telescope played in these
discoveries.
\end{abstract}

\section{Introduction}

Light from an astronomical source may reach an observer directly or
after being scattered by interstellar dust. In the latter case, the
arrival of any variations from the source object will be delayed by
the longer path length (relative to the direct path) - resulting in
what is referred to as a light echo.
\citep[e.g.,][]{Couderc39,Chevalier86,Schaefer87a,Xu94,Sugerman03,Patat05}.
The first light echoes were discovered around a nova in 1901, Nova
Persei, and shortly thereafter recognized as such
\citep{Ritchey01b,Kapteyn02,Perrine03}.  Since then, light echoes have
been observed around Cepheids \citep[e.g., RS
  Puppis,][]{Westerlund61}, eruptive variables \citep[e.g., V838
  Monocerotis,][]{Bond03}, young stellar objects \citep[e.g.,
  S~CrA,][]{Ortiz10}, and supernovae (SNe), both in the local group
\citep[e.g., SN~1987A,][]{Crotts88,Suntzeff88} and beyond \citep[e.g.,
  1991T,][and many more since]{Schmidt94}. A characteristic of all
these light echo instances is that they were discovered
serendipitously while the variable object or transient source was
still bright.

The idea of learning more about historical SNe by finding and studying
their scattered light echoes was first raised by \cite{Zwicky40}, but
the first dedicated surveys in the last century were not successful
\citep{vandenBergh65b,vandenBergh65a,vandenBergh66,Boffi99}.  However,
at the turn of the 21st century the development of focal planes populated
with large numbers of CCD detectors, in combination with advancements in
telescope technology, allowed the astronomical community to undertake
wide-field, time-domain surveys with unprecedented depth and area.
This lead to the first serendipitous discovery of light echoes of centuries-old SNe
in the Large Magellanic Cloud (LMC) by \cite{Rest05b} during the SuperMACHO 
survey \citep{Rest05a}.

Such light echoes of ancient events give us a rare opportunity in
astronomy: the direct observation of the cause (the
explosion/eruption) and the effect (the remnant) of the same
astronomical event. They also allow us to observe the explosion from
different angles, and thence to map asymmetries in the explosion. In
this paper, we discuss the crucial role of the Cerro Tololo
Interamerican Observatory (CTIO) Blanco telescope in the discovery and
follow-up of these light echoes.

\section{Light Echoes in the LMC\label{sec:lmc}}

\begin{figure}[b]
% \vspace*{-2.0 cm}
\begin{center}
 \includegraphics[width=3.4in]{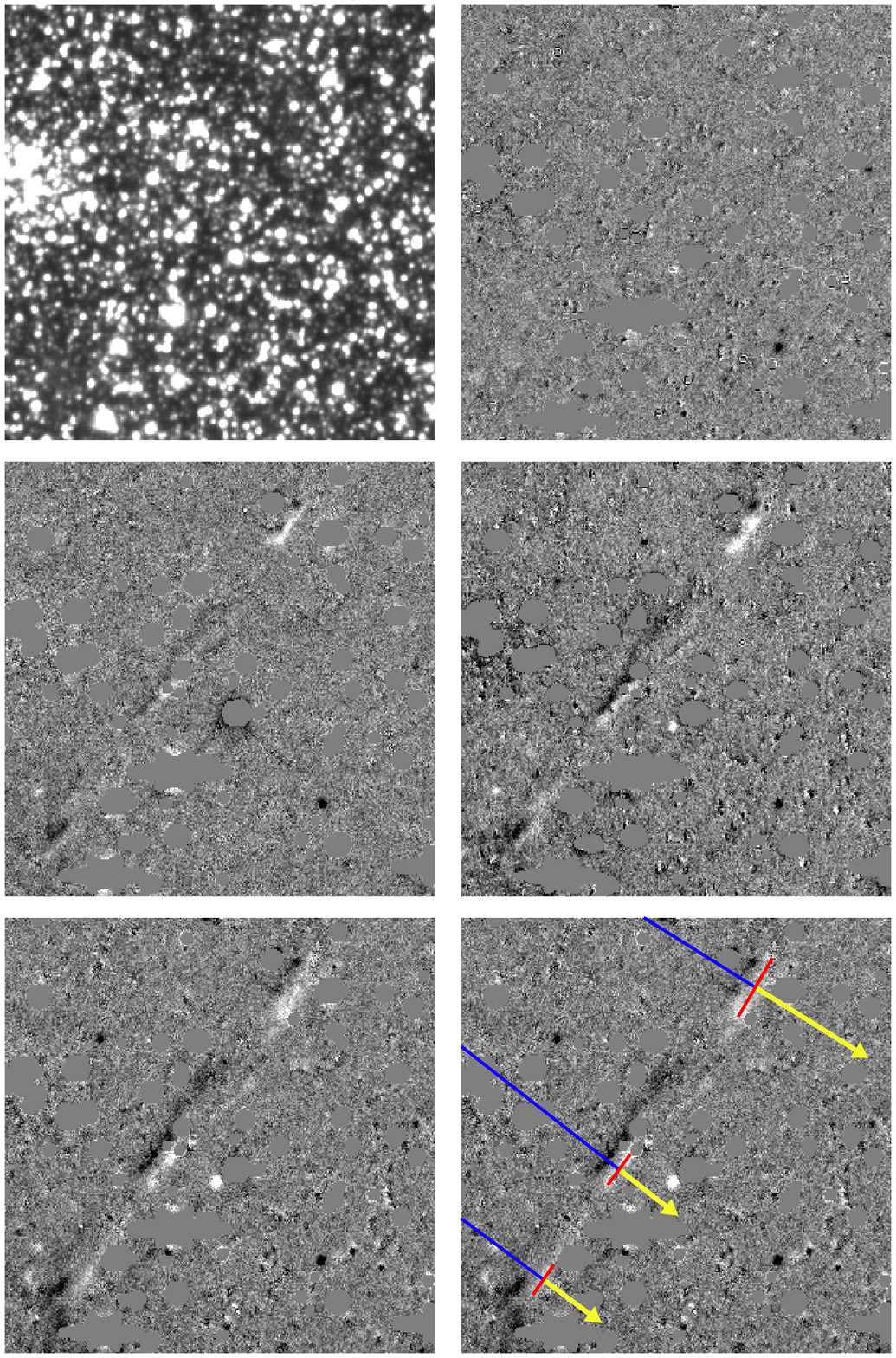} 
% \vspace*{-1.0 cm}
 \caption{Light echoes in the LMC from ancient SNe. Panel 1 (upper
left) shows the unsubtracted (template) Blanco CTIO 4m image which
includes the cluster Hodge 243.  Panel 2 (upper right) shows how
cleanly the field subtracts with data taken 50 days apart.  The next
three panels show the echo motion 1, 2, and 3 years after the template
date. White represents positive flux in the present epoch image and
black in the template image.  The vector motions are plotted in Panel
6 (lower right).  Each echo is fit with a straight line (red). The
apparent proper motion is given by the yellow vector and extrapolated
backwards (blue).  Saturated stars are masked out with grey circles. A
number of faint variable stars appear as black or white spots. This
figure and caption is courtesy of \cite{Rest05b}.}
\label{fig:lmcleexample}
\end{center}
\end{figure}

\begin{figure}[b]
% \vspace*{-2.0 cm}
\begin{center}
 \includegraphics[width=5in]{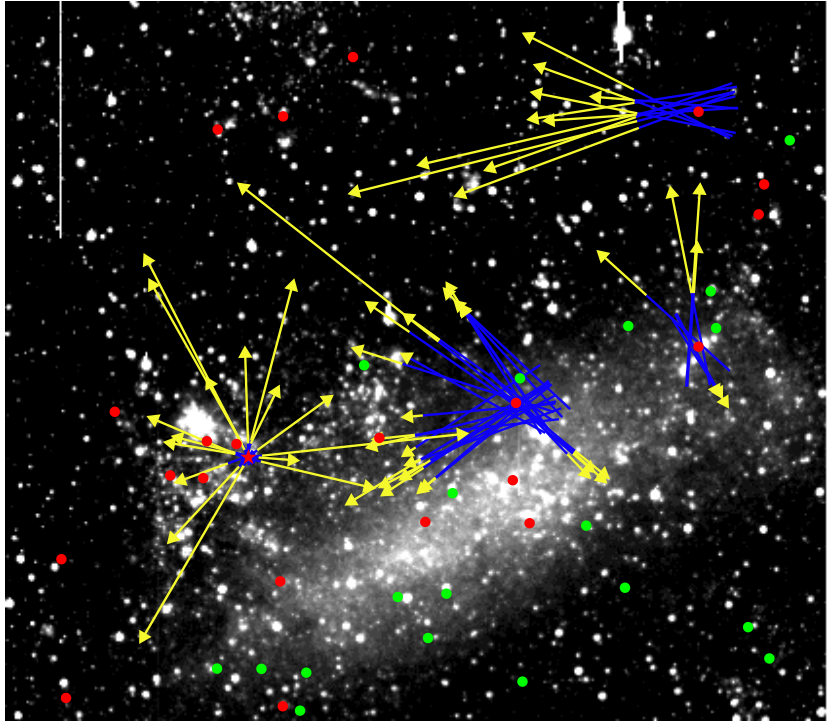} 
% \vspace*{-1.0 cm}
 \caption{A plot of the light echo vectors in the LMC. The vectors
have the same meaning as in Figure~\ref{fig:lmcleexample}. The centres of the echo
complexes are indicated by yellow circles. The source on the left
marked with a star is SN1987A.  The green circles are the location of
historical novae, and the red circles are the supernova remnant
locations. Evidently, the three light echo complexes point to
SNR 0519-69.0, SNR 0509-67.5, and SNR 0509-68.7.  This figure and caption is courtesy of
\cite{Rest05b}.}
\label{fig:lmcallle}
\end{center}
\end{figure}

The advent of CCDs and telescopes with large field-of-view allowed the
astronomical community to monitor large areas of the sky down to
(stellar) visual magnitudes of 23 and fainter.  The CTIO Blanco 4m
telescope hosted some of these first transient surveys.  One, the
microlensing survey SuperMACHO \citep{Rest05a}, monitored the LMC over
5 years. They reduced the Mosaic II CCD images with an automated
pipeline producing difference images and transient alerts, identifying
tens of thousands of transients \citep{Garg07}, variables
\citep{Garg10}, and the well-known light echoes of SN~1987A
\citep{Newman06, Sinnott13}.  Besides these expected variable sources,
the SuperMACHO Project also found extended features sharing
characteristics with the SN~1987A light echoes, but with significantly
slower apparent proper motions - a few arcsec~yr$^{-1}$ - and apparent
motion vector directions inconsistent with SN~1987A light echoes (see
Figure~\ref{fig:lmcleexample}).  In order to identify the sources of
these light echoes, if they were indeed light echoes, \cite{Rest05b}
overplotted the apparent motion vectors on an image of the LMC (see
Figure~\ref{fig:lmcallle}), and they found that the echo motions trace
back to three of its youngest supernova remnants (SNRs): SNR
0519-69.0, SNR 0509-67.5, and SNR 0509-68.7 (N103B).  These three
remnants had also been identified as Type Ia events, based on the
X-ray spectral abundances \citep{Hughes95}. Using the apparent motion
of their light echoes, the associated SNe were estimated to be between 400
and 800 years old \citep{Rest05b}.

\begin{figure}[b]
% \vspace*{-2.0 cm}
\begin{center}
 \includegraphics[width=5in]{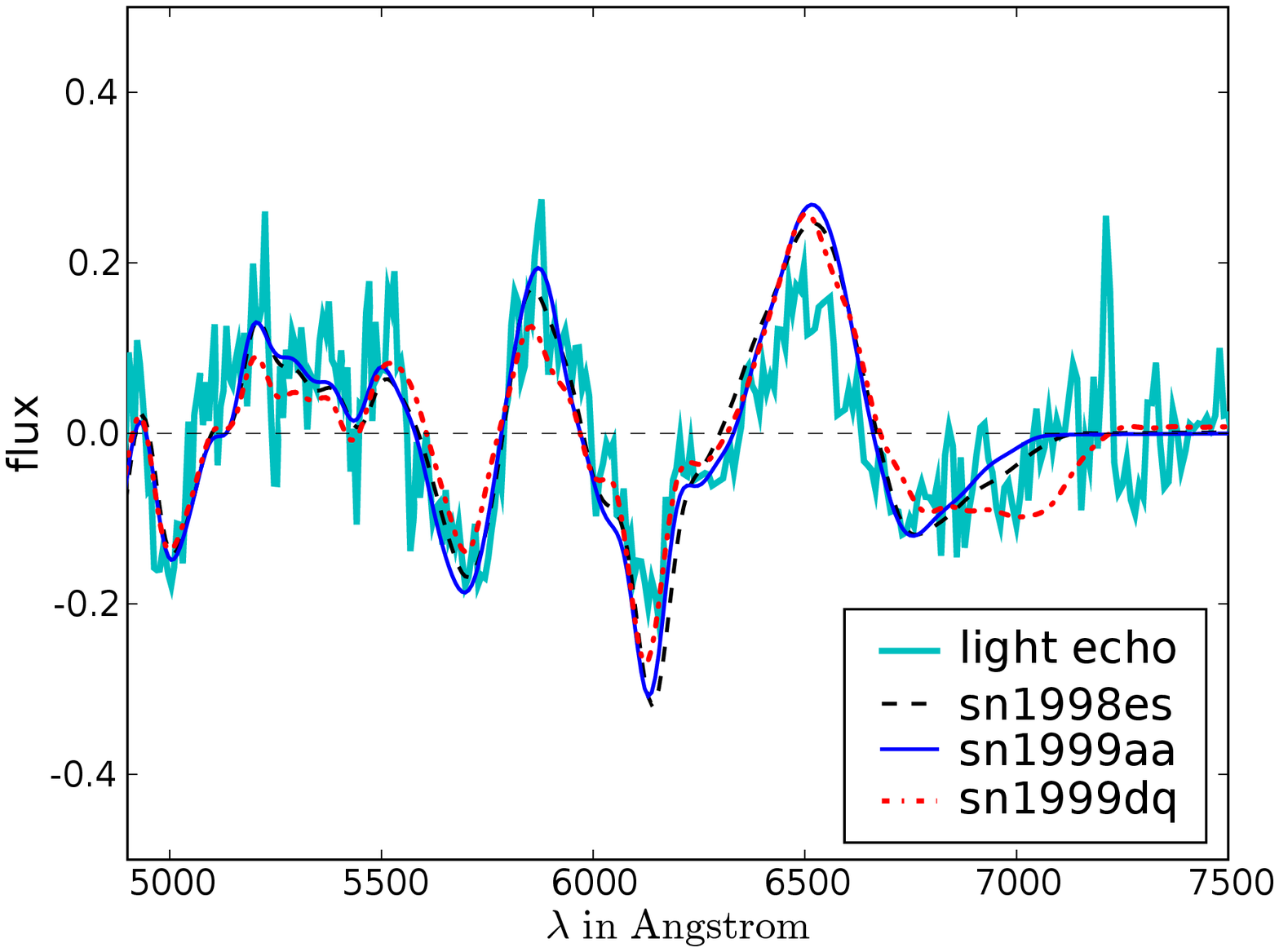} 
% \vspace*{-1.0 cm}
 \caption{The three time-integrated, dust-scattered, reddened, and
continuum-subtracted SN~Ia that best fit the light echo spetrum
associated with SNR 0519-69.0.  Note that all three template spectra
are 91T-like SNe~Ia with $\dm15<0.9$.  This figure and caption is
courtesy of \cite{Rest08a}.}
\label{fig:lmclespec}
\end{center}
\end{figure}

This was the first time that light echoes of ancient transients were
discovered. It allowed us to spectroscopically these transients long
after their light first reached Earth directly.  \cite{Rest08a}
analyzed a light echo from SNR 0509-675, which showed broad emission
and absorption lines consistent with a supernova spectrum.  To first
order, the observed light echo spectrum is the lightcurve-weighted
integration of the transients' individual spectral
epochs. \cite{Rest08a} compared the light echo spectrum to a spectral
library consisting of 26 SNe~Ia and 6 SN~Ib/c that were
time-integrated, corrected for the effect of being scattered by LMC
dust, and reddened by transit through the LMC and the Galaxy. When the
echo spectra where compared with the convolved SN spectra, they found
that overluminous 91T-like SNe~Ia with $\dm15<0.9$ (a light-curve
shape parameter where smaller indicates a more luminous SN) matched
the observed spectrum best (see Figure~\ref{fig:lmclespec}). An
analysis of SNR 0509-675 X-ray spectra is in excellent agreement with
this result \citep{Badenes08}.

The fact that light echoes of ancient SNe were found in the LMC
indicated that they also should be detectable within our own Milky Way
Galaxy, and it inspired survey programs to identify light echoes of
historic SNe in our own Galaxy. The challenge was to find the light
echoes, since the search radius scales with the inverse distance.
At this writing light echoes from two of the six historical supernovae,
Tycho (Ia) and Cas A (IIb), have been located and
spectroscopically classified \citep{Rest08b,Krause08a,Krause08b}.
In addition to the spectral typing, light echoes also offer two more
exciting scientific opportunities, 3D spectroscopy and spectroscopic
time series of transients, which are described in the the following
sections using SN~1987A and $\eta$~Carinae as examples.

\subsection{3D Spectroscopy of SN~1987A with Light Echoes\label{sec:87A}}

The appearance of SN 1987A in the Large Magellanic Cloud in 1987 was a
watershed event for many aspects of supernova astrophysics. Soon after
the outburst, it became apparent that there were significant amounts
of both circumstellar dust
\citep{Crotts91,Crotts95,Sugerman05a,Sugerman05b} and interstellar
dust \citep{Crotts88,Suntzeff88,Gouiffes88,Couch90,Xu94,Xu95} which
occupied the region around the supernova within the every-growing
light echo ellipsoid. At any given instant after the outburst, the
intersections of the locations of interstellar dust with the light
echo ellipsoid provides the observer with a set of distinct
perspectives on the outburst which differ from our usual direct,
line-of-sight perspective. By obtaining spectra of the outburst light
scattered from the instellar dust, it is possible to determine the
degree of asymmetry of the outburst. \cite{Boumis98} and
\cite{Smith03_eta} successfully applied this same technique to
characterize the luminous blue variable $\eta$ Car from spectra taken
of different locations in its surrounding reflection
nebula. \cite{Rest11_casaspec} used this technique with light echoes
of Cas~A --- the first time that this technique has been applied to a
SN.

\begin{figure}[b]
% \vspace*{-2.0 cm}
\begin{center}
 \includegraphics[width=3.4in]{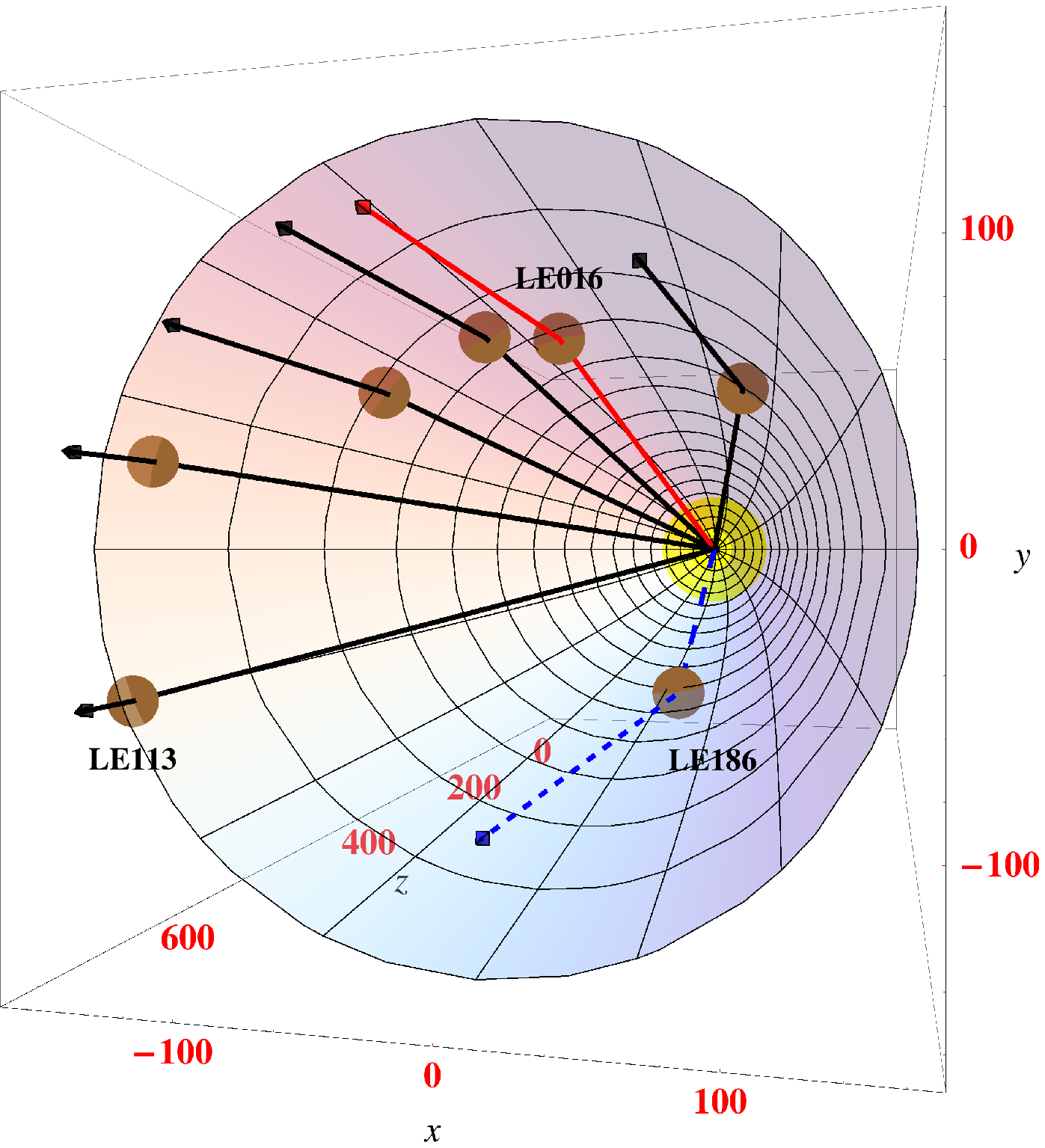} 
% \vspace*{-1.0 cm}
 \caption{3D locations of the scattering dust for the seven light echoes of
SN~1987A that were spectroscopically followed up. Highlighted in solid
red and dashed blue are the extreme viewing angles corresponding to
LE016 and LE186. North is towards the positive y axis, east is towards
the negative x axis, and z is the distance in front of the SN. All
units are in light years. This figure and caption is courtesy of \cite{Sinnott13}.}
   \label{fig:87a3D}
\end{center}
\end{figure}

\begin{figure}[b]
\begin{center}
 \includegraphics[width=3.4in]{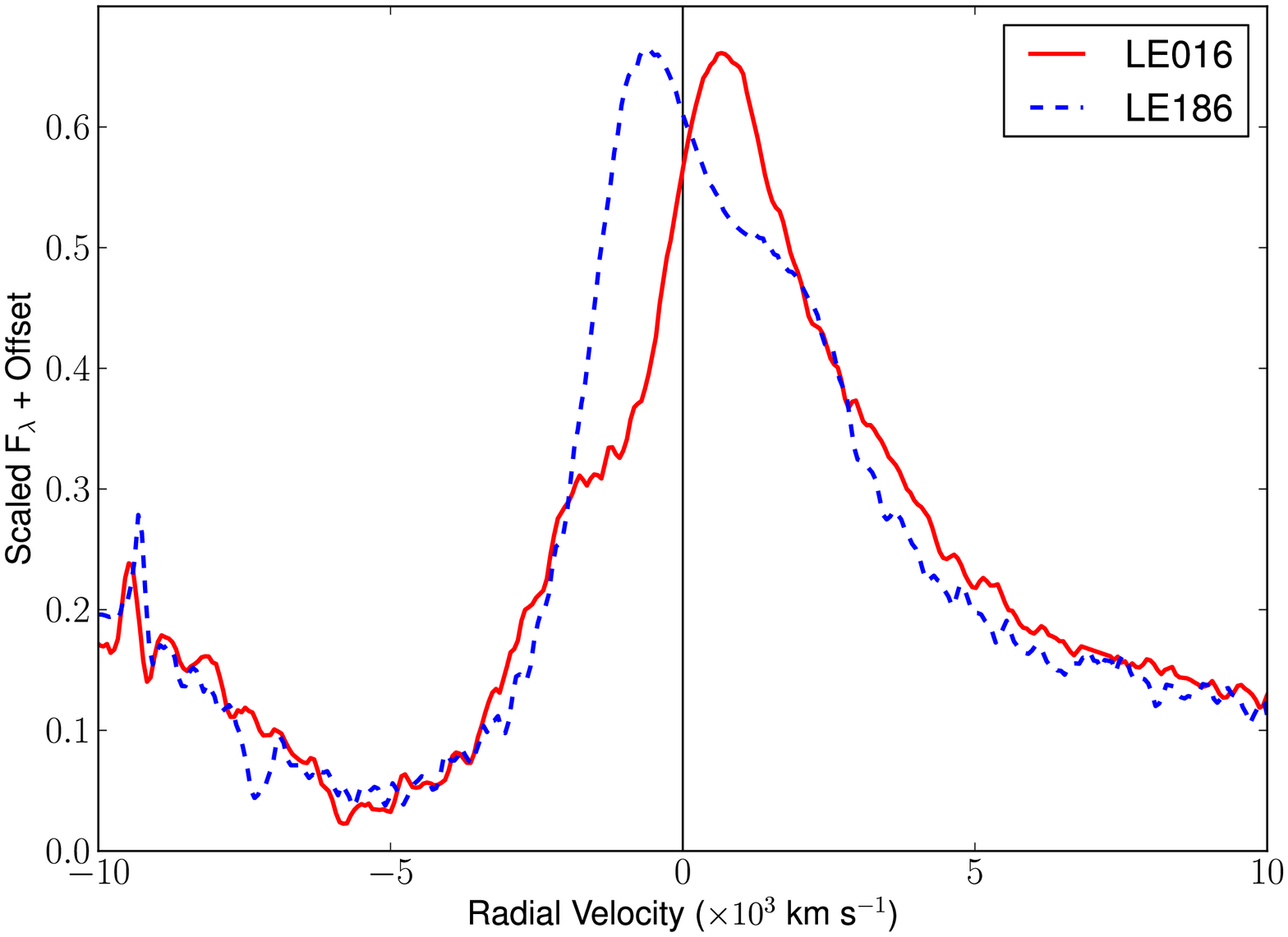} 
 \caption{Observed H$\alpha$ lines from LE016 and LE186. Emission peaks
have been interpolated with high-order polynomials. Spectra are scaled and
offset for comparison purposes, as well as smoothed with a boxcar of 3 pixels.
Although this plot does not take into account the important differences
in light echo time-integrations between the spectra, it highlights the overall difference
in fine-structure in the two light echo spectra from opposite PAs. Observing
H$\alpha$ profiles with opposite asymmetry structure at opposite PAs is surprising
considering the opening angle between the two light echoes is $<40\arcdeg$. 
This figure and caption is courtesy of \cite{Sinnott13}.}
   \label{fig:87aLEspec}
\end{center}
\end{figure}

Fig.~\ref{fig:87a3D} illustrates the collection of seven non-direct
perspectives from which spectra were obtained and reported by \cite{Sinnott13} 
to look for spectral line changes which might arise from asymmetry.
In that work, the spectra of the light echo locations was corrected
for the dust's enhanced scattering at shorter wavelengths and then
compared with a temporal library of spectrophotometry obtained at SAAO.
\cite{Sinnott13} found a excess in the H$\alpha$ P Cygni profile in the
red-shifted emission and a blue-shifted ``knee'' at PA=16\arcdeg
for the light echo labelled LE016 - (see the red line Figure~\ref{fig:87aLEspec}).

There is a clear pattern in the how the H$\alpha$ features change as a
function of position angle. The nearly diametrically-opposite light echo at
PA=186\arcdeg has a complementary pattern of an excess in blueshifted
emission and a redshifted ``knee'' - (see blue line in Figure~\ref{fig:87aLEspec})
Despite the maximum perspective different being only \about 40\arcdeg,
the easily-detected spectral profile changes suggest very significant
differences in the photosphere as seen from LE016 and LE186. \citet{Sinnott13}
found that a bipolar and asymmetric $^{56}$Ni distribution with a
symmetry axis aligned along positional angles 16\arcdeg/186\arcdeg
would provide a scenario consistent with their light echo spectra and the symmetry
axis (\about PA=15\arcdeg) of elongated ejecta measured by \cite{Wang02} and \cite{Kjaer10}.
The sign of the radial velocities of ejecta measured by \cite{Kjaer10} are
also consistent with the findings of \citet{Sinnott13}, tying the
asymmetry in the photosphere at the time of the outburst to the
emerging supernova remnant structure.

\subsection{Spectroscopic Time Series of $\eta$~Car's Great Eruption with Light Echoes\label{sec:etacar}}

In the idealized case where a scattering dust filament is infinitely
thin, the intensity of a light echo is simply the projected light
curve of the source transient.  Thus spectroscopy of a single light
echo can, in theory, provide spectra from individual epochs of the
transient.  However, the finite thickness of the scattering dust
filaments and the effects of seeing cause the convolution of nearby
epochs of the light echo profile into the observed spectrum
\citep{Rest11_leprofile,Rest12b_lerev}.  In the case of supernovae in
our Galaxy and typical dust filaments, the temporal resolution is
typically between 3 weeks to a couple of months. Only under the most
favorable conditions (small PSF size with space-based observations and
favorable dust thickness/inclination) is the convolution effect
reduced to as short as 1 week \citep{Rest11_leprofile}.  However, for
transients with longer time scales than SNe it is possible to resolve
them even using ground-based spectroscopy. A good example of this
latter situation is the Great Eruption of $\eta$~Car, which lasted for
two decades during the nineteenth century and showed temporal
variability on time-scales of months and years.

\begin{figure}[b]
% \vspace*{-2.0 cm}
\begin{center}
 \includegraphics[width=2.4in]{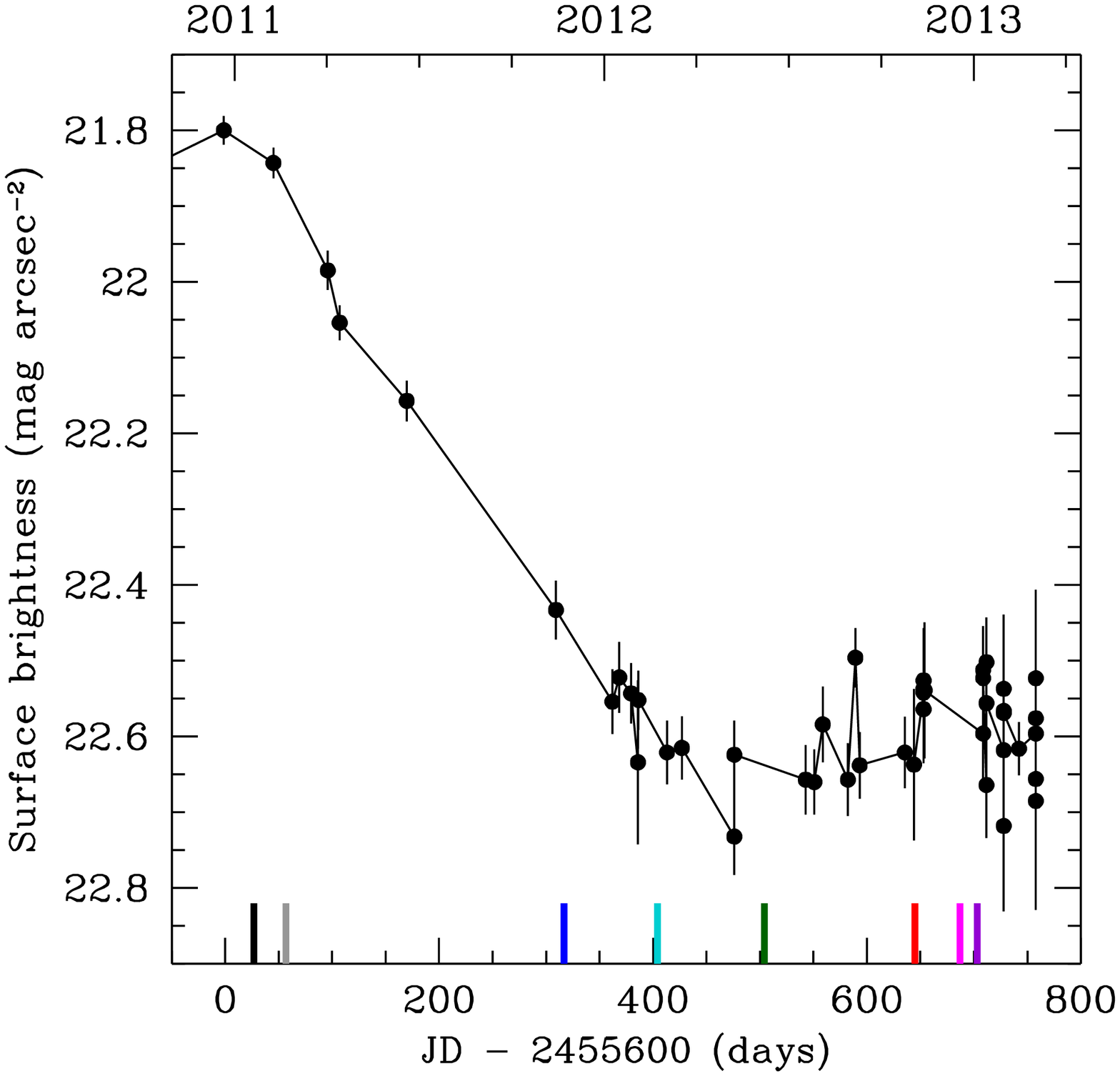} 
 \includegraphics[width=2.4in]{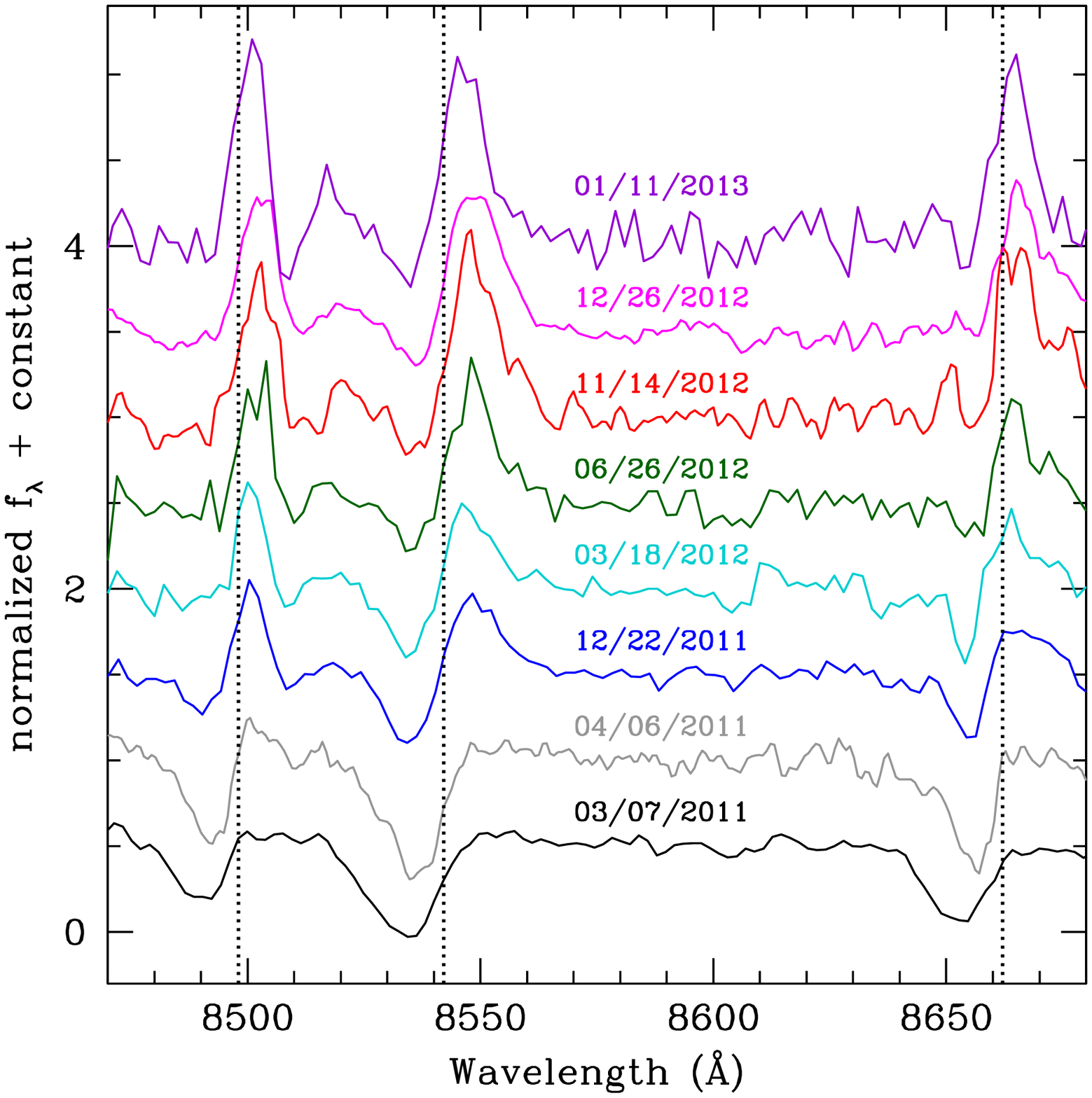} 
% \vspace*{-1.0 cm}
 \caption{{\it Left:} Light curve of a portion of $\eta$~Car's Great Eruption,
derived from its light echoes. Shown is the surface brightness in $i$ band from
the same sky position at different epochs (Blanco 4m (MOSAIC II,
DECam), Swope (Direct CCD), FTS (Spectral), SOAR (SOI)). The light echo is the
projected light curve of the source transient. Since the light echo
has an apparent motion, the light curve of the source event ``moves''
through a given position on the sky. The epochs spectra were taken are
indicated with the colored lines.  {\it Right:} LE spectra of $\eta$~Car's
Great Eruption showing the Ca~II IR triplet from different epochs at
the same sky location (Magellan Baade (IMACS), Gemini-S (GMOS)).
The epochs of the spectra are indicated with in the left panel 
with vertical bars of the same
color as the spectra.
} 
  \label{fig:EClcCaII}
\end{center}
\end{figure}

\begin{figure}[b]
% \vspace*{-2.0 cm}
\begin{center}
 \includegraphics[width=3.4in]{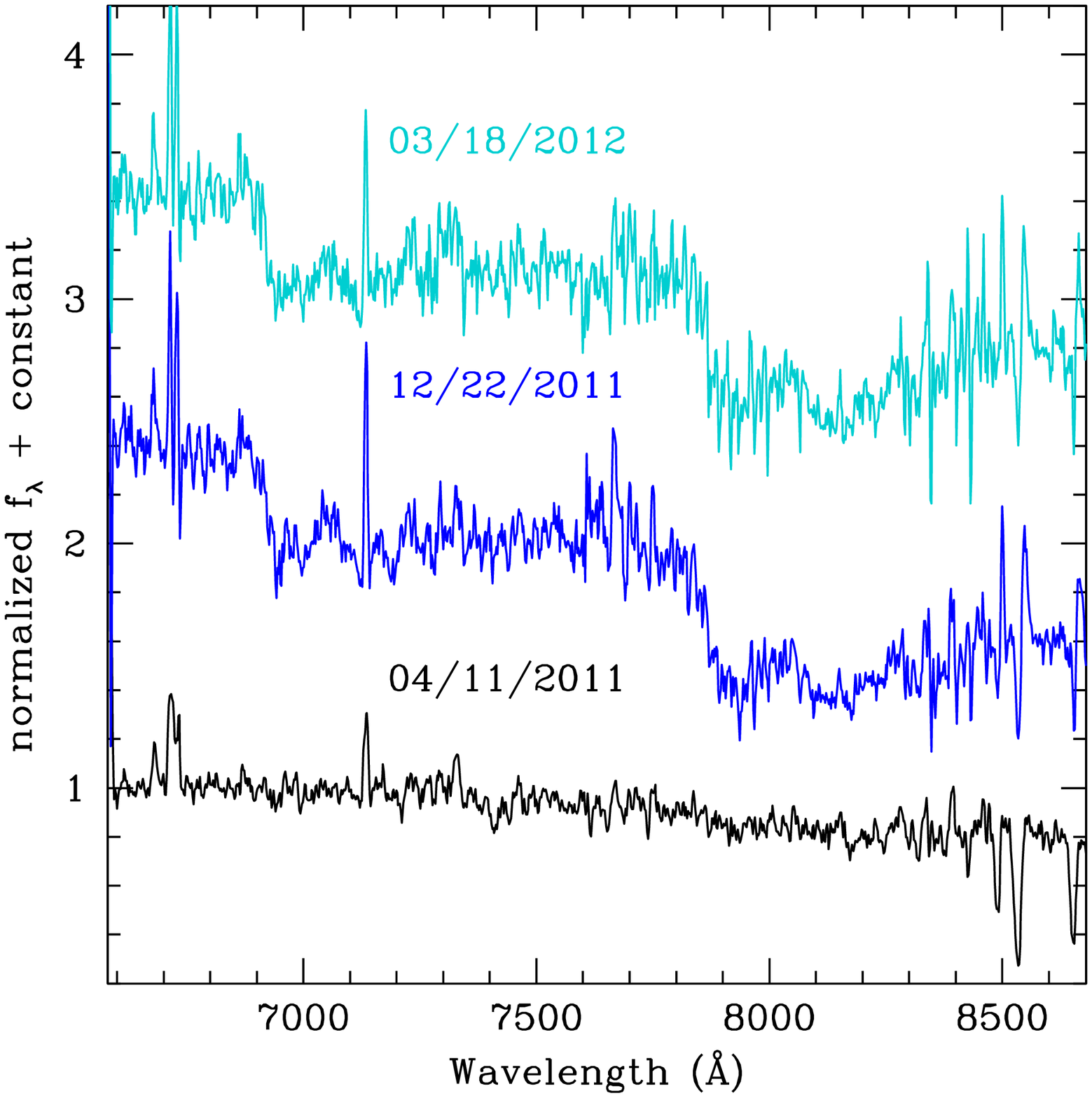} 
% \vspace*{-1.0 cm}
 \caption{LE spectra of $\eta$~Car's Great Eruption showing how the CN
   molecular bands develop for later epochs.}
   \label{fig:CN}
\end{center}
\end{figure}

The left panel of Figure~\ref{fig:EClcCaII} shows the flux from a
light echo of $\eta$~Car's Great Eruption at the same location for
various epochs: the observed flux time series is the light curve of
the Great Eruption, convolved with the scattering dust thickness and
image point spread function.  The colored vertical lines indicate the
epochs at which we obtained spectra of this light echo at that
position, shown in the same colors in the right panel.  This single
light echo allowed us to analyze the evolution of the Ca~II lines
during $\eta$~Car's Great Eruption, and we found distinct differences
from the typical spectral evolution of LBVs. The earliest spectra
taken at epochs close to the peak in the light curve correlate best
with spectra of G2-to-G5 supergiants, a later spectral type than predicted by
standard opaque wind models \citep{Rest12_etac}.  The early Ca~II IR
triplet features are in absorption only, with an average blueshift of
$-200$~\kms, and the lines are asymmetric extending up to $-800$~\kms
\citep{Rest12_etac}. Later spectra, corresponding to the
decline phase of the burst, show that the Ca~II IR triplet evolved from
pure absorption, through a P-Cygni profile, to a nearly pure emission
line spectrum.  That transformation was accompanied by the development
of strong CN molecular bands (see Figure~\ref{fig:CN}).  In contrast,
standard LBV outburst spectra return to the earlier stellar
types toward the end of an eruption. A paper detailing these observations 
is in preparation.

\acknowledgements 

We thank all the observers that have contributed to the monitoring of
$\eta$~Car's light echoes, especially E.~Hsiao (and the Carnegie
Supernova Project II).  Based on observations of program GS-2012B-Q-57
obtained at the Gemini Observatory, which is operated by the
Association of Universities for Research in Astronomy, Inc., under a
cooperative agreement with the NSF on behalf of the Gemini
partnership: the National Science Foundation (United States), the
National Research Council (Canada), CONICYT (Chile), the Australian
Research Council (Australia), Minist\'{e}rio da Ci\^{e}ncia,
Tecnologia e Inova\c{c}\~{a}o (Brazil) and Ministerio de Ciencia,
Tecnolog\'{i}a e Innovaci\'{o}n Productiva (Argentina).  This paper
includes data gathered with the 6.5 meter Magellan Telescopes located
at Las Campanas Observatory, Chile. Based on observations at the Cerro
Tololo Inter-American Observatory, National Optical Astronomy
Observatory, which are operated by the Association of Universities for
Research in Astronomy, under contract with the National Science
Foundation.  The SOAR Telescope is a joint project of: Conselho
Nacional de Pesquisas Cientificas e Tecnologicas CNPq-Brazil, The
University of North Carolina at Chapel Hill, Michigan State
University, and the National Optical Astronomy Observatory. The
Faulkes Telescope South (FTS) is operated and maintained by the Las
Cumbres Observatory Global Telescope. This work was supported by the HST programs GO-12577 and  AR-12851.

\bibliography{rest}

\begin{thebibliography}{}
\expandafter\ifx\csname natexlab\endcsname\relax\def\natexlab#1{#1}\fi
\expandafter\ifx\csname url\endcsname\relax
  \def\url#1{\texttt{#1}}\fi
\expandafter\ifx\csname urlprefix\endcsname\relax\def\urlprefix{URL }\fi
\providecommand{\eprint}[2][]{\url{#2}}

\bibitem[{{Badenes} et~al.(2008){Badenes}, {Hughes}, {Cassam-Chena{\"i}}, \&
  {Bravo}}]{Badenes08}
{Badenes}, C., {Hughes}, J.~P., {Cassam-Chena{\"i}}, G., \& {Bravo}, E. 2008,
  \apj, 680, 1149. \eprint{0801.4761}

\bibitem[{{Boffi} et~al.(1999){Boffi}, {Sparks}, \& {Macchetto}}]{Boffi99}
{Boffi}, F.~R., {Sparks}, W.~B., \& {Macchetto}, F.~D. 1999, \aaps, 138, 253.
  \eprint{astro-ph/9906206}

\bibitem[{{Bond} et~al.(2003){Bond}, {Henden}, {Levay}, {Panagia}, {Sparks},
  {Starrfield}, {Wagner}, {Corradi}, \& {Munari}}]{Bond03}
{Bond}, H.~E., {Henden}, A., {Levay}, Z.~G., {Panagia}, N., {Sparks}, W.~B.,
  {Starrfield}, S., {Wagner}, R.~M., {Corradi}, R.~L.~M., \& {Munari}, U. 2003,
  \nat, 422, 405. \eprint{astro-ph/0303513}

\bibitem[{{Boumis} et~al.(1998){Boumis}, {Meaburn}, {Bryce}, \&
  {Lopez}}]{Boumis98}
{Boumis}, P., {Meaburn}, J., {Bryce}, M., \& {Lopez}, J.~A. 1998, \mnras, 294,
  61

\bibitem[{{Chevalier}(1986)}]{Chevalier86}
{Chevalier}, R.~A. 1986, \apj, 308, 225

\bibitem[{{Couch} et~al.(1990){Couch}, {Allen}, \& {Malin}}]{Couch90}
{Couch}, W.~J., {Allen}, D.~A., \& {Malin}, D.~F. 1990, \mnras, 242, 555

\bibitem[{{Couderc}(1939)}]{Couderc39}
{Couderc}, P. 1939, Annales d'Astrophysique, 2, 271

\bibitem[{{Crotts}(1988)}]{Crotts88}
{Crotts}, A. 1988, \iaucirc, 4561, 4

\bibitem[{{Crotts} \& {Kunkel}(1991)}]{Crotts91}
{Crotts}, A.~P.~S., \& {Kunkel}, W.~E. 1991, \apjl, 366, L73

\bibitem[{{Crotts} et~al.(1995){Crotts}, {Kunkel}, \& {Heathcote}}]{Crotts95}
{Crotts}, A.~P.~S., {Kunkel}, W.~E., \& {Heathcote}, S.~R. 1995, \apj, 438, 724

\bibitem[{{Garg} et~al.(2010){Garg}, {Cook}, {Nikolaev}, {Huber}, {Rest},
  {Becker}, {Challis}, {Clocchiatti}, {Miknaitis}, {Minniti}, {Morelli},
  {Olsen}, {Prieto}, {Suntzeff}, {Welch}, \& {Wood-Vasey}}]{Garg10}
{Garg}, A., {Cook}, K.~H., {Nikolaev}, S., {Huber}, M.~E., {Rest}, A.,
  {Becker}, A.~C., {Challis}, P., {Clocchiatti}, A., {Miknaitis}, G.,
  {Minniti}, D., {Morelli}, L., {Olsen}, K., {Prieto}, J.~L., {Suntzeff},
  N.~B., {Welch}, D.~L., \& {Wood-Vasey}, W.~M. 2010, \aj, 140, 328.
  \eprint{1004.0955}

\bibitem[{{Garg} et~al.(2007){Garg}, {Stubbs}, {Challis}, {Wood-Vasey},
  {Blondin}, {Huber}, {Cook}, {Nikolaev}, {Rest}, {Smith}, {Olsen}, {Suntzeff},
  {Aguilera}, {Prieto}, {Becker}, {Miceli}, {Miknaitis}, {Clocchiatti},
  {Minniti}, {Morelli}, \& {Welch}}]{Garg07}
{Garg}, A., {Stubbs}, C.~W., {Challis}, P., {Wood-Vasey}, W.~M., {Blondin}, S.,
  {Huber}, M.~E., {Cook}, K., {Nikolaev}, S., {Rest}, A., {Smith}, R.~C.,
  {Olsen}, K., {Suntzeff}, N.~B., {Aguilera}, C., {Prieto}, J.~L., {Becker},
  A., {Miceli}, A., {Miknaitis}, G., {Clocchiatti}, A., {Minniti}, D.,
  {Morelli}, L., \& {Welch}, D.~L. 2007, \aj, 133, 403.
  \eprint{arXiv:astro-ph/0608639}

\bibitem[{{Gouiffes} et~al.(1988){Gouiffes}, {Rosa}, {Melnick}, {Danziger},
  {Remy}, {Santini}, {Sauvageot}, {Jakobsen}, \& {Ruiz}}]{Gouiffes88}
{Gouiffes}, C., {Rosa}, M., {Melnick}, J., {Danziger}, I.~J., {Remy}, M.,
  {Santini}, C., {Sauvageot}, J.~L., {Jakobsen}, P., \& {Ruiz}, M.~T. 1988,
  \aap, 198, L9

\bibitem[{{Hughes} et~al.(1995){Hughes}, {Hayashi}, {Helfand}, {Hwang}, {Itoh},
  {Kirshner}, {Koyama}, {Markert}, {Tsunemi}, \& {Woo}}]{Hughes95}
{Hughes}, J.~P., {Hayashi}, I., {Helfand}, D., {Hwang}, U., {Itoh}, M.,
  {Kirshner}, R., {Koyama}, K., {Markert}, T., {Tsunemi}, H., \& {Woo}, J.
  1995, \apjl, 444, L81

\bibitem[{{Kapteyn}(1902)}]{Kapteyn02}
{Kapteyn}, J.~C. 1902, Astronomische Nachrichten, 157, 201

\bibitem[{{Kj{\ae}r} et~al.(2010){Kj{\ae}r}, {Leibundgut}, {Fransson},
  {Jerkstrand}, \& {Spyromilio}}]{Kjaer10}
{Kj{\ae}r}, K., {Leibundgut}, B., {Fransson}, C., {Jerkstrand}, A., \&
  {Spyromilio}, J. 2010, \aap, 517, A51. \eprint{1003.5684}

\bibitem[{{Krause} et~al.(2008{\natexlab{a}}){Krause}, {Birkmann}, {Usuda},
  {Hattori}, {Goto}, {Rieke}, \& {Misselt}}]{Krause08a}
{Krause}, O., {Birkmann}, S.~M., {Usuda}, T., {Hattori}, T., {Goto}, M.,
  {Rieke}, G.~H., \& {Misselt}, K.~A. 2008{\natexlab{a}}, Science, 320, 1195.
  \eprint{0805.4557}

\bibitem[{{Krause} et~al.(2008{\natexlab{b}}){Krause}, {Tanaka}, {Usuda},
  {Hattori}, {Goto}, {Birkmann}, \& {Nomoto}}]{Krause08b}
{Krause}, O., {Tanaka}, M., {Usuda}, T., {Hattori}, T., {Goto}, M., {Birkmann},
  S., \& {Nomoto}, K. 2008{\natexlab{b}}, \nat, 456, 617. \eprint{0810.5106}

\bibitem[{{Newman} \& {Rest}(2006)}]{Newman06}
{Newman}, A.~B., \& {Rest}, A. 2006, \pasp, 118, 1484.
  \eprint{arXiv:astro-ph/0610579}

\bibitem[{{Ortiz} et~al.(2010){Ortiz}, {Sugerman}, {de La Cueva},
  {Santos-Sanz}, {Duffard}, {Gil-Hutton}, {Melita}, \& {Morales}}]{Ortiz10}
{Ortiz}, J.~L., {Sugerman}, B.~E.~K., {de La Cueva}, I., {Santos-Sanz}, P.,
  {Duffard}, R., {Gil-Hutton}, R., {Melita}, M., \& {Morales}, N. 2010, \aap,
  519, A7+. \eprint{1007.2556}

\bibitem[{{Patat}(2005)}]{Patat05}
{Patat}, F. 2005, \mnras, 357, 1161. \eprint{astro-ph/0409666}

\bibitem[{{Perrine}(1903)}]{Perrine03}
{Perrine}, C.~D. 1903, \apj, 17, 310

\bibitem[{{Rest} et~al.(2011{\natexlab{a}}){Rest}, {Foley}, {Sinnott}, {Welch},
  {Badenes}, {Filippenko}, {Bergmann}, {Bhatti}, {Blondin}, {Challis}, {Damke},
  {Finley}, {Huber}, {Kasen}, {Kirshner}, {Matheson}, {Mazzali}, {Minniti},
  {Nakajima}, {Narayan}, {Olsen}, {Sauer}, {Smith}, \&
  {Suntzeff}}]{Rest11_casaspec}
{Rest}, A., {Foley}, R.~J., {Sinnott}, B., {Welch}, D.~L., {Badenes}, C.,
  {Filippenko}, A.~V., {Bergmann}, M., {Bhatti}, W.~A., {Blondin}, S.,
  {Challis}, P., {Damke}, G., {Finley}, H., {Huber}, M.~E., {Kasen}, D.,
  {Kirshner}, R.~P., {Matheson}, T., {Mazzali}, P., {Minniti}, D., {Nakajima},
  R., {Narayan}, G., {Olsen}, K., {Sauer}, D., {Smith}, R.~C., \& {Suntzeff},
  N.~B. 2011{\natexlab{a}}, \apj, 732, 3

\bibitem[{{Rest} et~al.(2008{\natexlab{a}}){Rest}, {Matheson}, {Blondin},
  {Bergmann}, {Welch}, {Suntzeff}, {Smith}, {Olsen}, {Prieto}, {Garg},
  {Challis}, {Stubbs}, {Hicken}, {Modjaz}, {Wood-Vasey}, {Zenteno}, {Damke},
  {Newman}, {Huber}, {Cook}, {Nikolaev}, {Becker}, {Miceli}, {Covarrubias},
  {Morelli}, {Pignata}, {Clocchiatti}, {Minniti}, \& {Foley}}]{Rest08a}
{Rest}, A., {Matheson}, T., {Blondin}, S., {Bergmann}, M., {Welch}, D.~L.,
  {Suntzeff}, N.~B., {Smith}, R.~C., {Olsen}, K., {Prieto}, J.~L., {Garg}, A.,
  {Challis}, P., {Stubbs}, C., {Hicken}, M., {Modjaz}, M., {Wood-Vasey}, W.~M.,
  {Zenteno}, A., {Damke}, G., {Newman}, A., {Huber}, M., {Cook}, K.~H.,
  {Nikolaev}, S., {Becker}, A.~C., {Miceli}, A., {Covarrubias}, R., {Morelli},
  L., {Pignata}, G., {Clocchiatti}, A., {Minniti}, D., \& {Foley}, R.~J.
  2008{\natexlab{a}}, \apj, 680, 1137. \eprint{0801.4762}

\bibitem[{{Rest} et~al.(2012{\natexlab{a}}){Rest}, {Prieto}, {Walborn},
  {Smith}, {Bianco}, {Chornock}, {Welch}, {Howell}, {Huber}, {Foley}, {Fong},
  {Sinnott}, {Bond}, {Smith}, {Toledo}, {Minniti}, \& {Mandel}}]{Rest12_etac}
{Rest}, A., {Prieto}, J.~L., {Walborn}, N.~R., {Smith}, N., {Bianco}, F.~B.,
  {Chornock}, R., {Welch}, D.~L., {Howell}, D.~A., {Huber}, M.~E., {Foley},
  R.~J., {Fong}, W., {Sinnott}, B., {Bond}, H.~E., {Smith}, R.~C., {Toledo},
  I., {Minniti}, D., \& {Mandel}, K. 2012{\natexlab{a}}, \nat, 482, 375.
  \eprint{1112.2210}

\bibitem[{{Rest} et~al.(2012{\natexlab{b}}){Rest}, {Sinnott}, \&
  {Welch}}]{Rest12b_lerev}
{Rest}, A., {Sinnott}, B., \& {Welch}, D.~L. 2012{\natexlab{b}}, \pasa, 29,
  466. \eprint{1204.1341}

\bibitem[{{Rest} et~al.(2011{\natexlab{b}}){Rest}, {Sinnott}, {Welch}, {Foley},
  {Narayan}, {Mandel}, {Huber}, \& {Blondin}}]{Rest11_leprofile}
{Rest}, A., {Sinnott}, B., {Welch}, D.~L., {Foley}, R.~J., {Narayan}, G.,
  {Mandel}, K., {Huber}, M.~E., \& {Blondin}, S. 2011{\natexlab{b}}, \apj, 732,
  2

\bibitem[{{Rest} et~al.(2005{\natexlab{a}}){Rest}, {Stubbs}, {Becker},
  {Miknaitis}, {Miceli}, {Covarrubias}, {Hawley}, {Smith}, {Suntzeff}, {Olsen},
  {Prieto}, {Hiriart}, {Welch}, {Cook}, {Nikolaev}, {Huber}, {Prochtor},
  {Clocchiatti}, {Minniti}, {Garg}, {Challis}, {Keller}, \&
  {Schmidt}}]{Rest05a}
{Rest}, A., {Stubbs}, C., {Becker}, A.~C., {Miknaitis}, G.~A., {Miceli}, A.,
  {Covarrubias}, R., {Hawley}, S.~L., {Smith}, R.~C., {Suntzeff}, N.~B.,
  {Olsen}, K., {Prieto}, J.~L., {Hiriart}, R., {Welch}, D.~L., {Cook}, K.~H.,
  {Nikolaev}, S., {Huber}, M., {Prochtor}, G., {Clocchiatti}, A., {Minniti},
  D., {Garg}, A., {Challis}, P., {Keller}, S.~C., \& {Schmidt}, B.~P.
  2005{\natexlab{a}}, \apj, 634, 1103

\bibitem[{{Rest} et~al.(2005{\natexlab{b}}){Rest}, {Suntzeff}, {Olsen},
  {Prieto}, {Smith}, {Welch}, {Becker}, {Bergmann}, {Clocchiatti}, {Cook},
  {Garg}, {Huber}, {Miknaitis}, {Minniti}, {Nikolaev}, \& {Stubbs}}]{Rest05b}
{Rest}, A., {Suntzeff}, N.~B., {Olsen}, K., {Prieto}, J.~L., {Smith}, R.~C.,
  {Welch}, D.~L., {Becker}, A., {Bergmann}, M., {Clocchiatti}, A., {Cook}, K.,
  {Garg}, A., {Huber}, M., {Miknaitis}, G., {Minniti}, D., {Nikolaev}, S., \&
  {Stubbs}, C. 2005{\natexlab{b}}, \nat, 438, 1132

\bibitem[{{Rest} et~al.(2008{\natexlab{b}}){Rest}, {Welch}, {Suntzeff},
  {Oaster}, {Lanning}, {Olsen}, {Smith}, {Becker}, {Bergmann}, {Challis},
  {Clocchiatti}, {Cook}, {Damke}, {Garg}, {Huber}, {Matheson}, {Minniti},
  {Prieto}, \& {Wood-Vasey}}]{Rest08b}
{Rest}, A., {Welch}, D.~L., {Suntzeff}, N.~B., {Oaster}, L., {Lanning}, H.,
  {Olsen}, K., {Smith}, R.~C., {Becker}, A.~C., {Bergmann}, M., {Challis}, P.,
  {Clocchiatti}, A., {Cook}, K.~H., {Damke}, G., {Garg}, A., {Huber}, M.~E.,
  {Matheson}, T., {Minniti}, D., {Prieto}, J.~L., \& {Wood-Vasey}, W.~M.
  2008{\natexlab{b}}, \apjl, 681, L81. \eprint{0805.4607}

\bibitem[{{Ritchey}(1901)}]{Ritchey01b}
{Ritchey}, G.~W. 1901, \apj, 14, 293

\bibitem[{{Schaefer}(1987)}]{Schaefer87a}
{Schaefer}, B.~E. 1987, \apjl, 323, L47

\bibitem[{{Schmidt} et~al.(1994){Schmidt}, {Kirshner}, {Leibundgut}, {Wells},
  {Porter}, {Ruiz-Lapuente}, {Challis}, \& {Filippenko}}]{Schmidt94}
{Schmidt}, B.~P., {Kirshner}, R.~P., {Leibundgut}, B., {Wells}, L.~A.,
  {Porter}, A.~C., {Ruiz-Lapuente}, P., {Challis}, P., \& {Filippenko}, A.~V.
  1994, \apjl, 434, L19. \eprint{astro-ph/9407097}

\bibitem[{{Sinnott} et~al.(2013){Sinnott}, {Welch}, {Rest}, {Sutherland}, \&
  {Bergmann}}]{Sinnott13}
{Sinnott}, B., {Welch}, D.~L., {Rest}, A., {Sutherland}, P.~G., \& {Bergmann},
  M. 2013, \apj, 767, 45. \eprint{1211.3781}

\bibitem[{{Smith} et~al.(2003){Smith}, {Davidson}, {Gull}, {Ishibashi}, \&
  {Hillier}}]{Smith03_eta}
{Smith}, N., {Davidson}, K., {Gull}, T.~R., {Ishibashi}, K., \& {Hillier},
  D.~J. 2003, \apj, 586, 432. \eprint{arXiv:astro-ph/0301394}

\bibitem[{{Sugerman}(2003)}]{Sugerman03}
{Sugerman}, B.~E.~K. 2003, \aj, 126, 1939. \eprint{astro-ph/0307245}

\bibitem[{{Sugerman} et~al.(2005{\natexlab{a}}){Sugerman}, {Crotts}, {Kunkel},
  {Heathcote}, \& {Lawrence}}]{Sugerman05a}
{Sugerman}, B.~E.~K., {Crotts}, A.~P.~S., {Kunkel}, W.~E., {Heathcote}, S.~R.,
  \& {Lawrence}, S.~S. 2005{\natexlab{a}}, \apj, 627, 888.
  \eprint{arXiv:astro-ph/0502268}

\bibitem[{{Sugerman} et~al.(2005{\natexlab{b}}){Sugerman}, {Crotts}, {Kunkel},
  {Heathcote}, \& {Lawrence}}]{Sugerman05b}
--- 2005{\natexlab{b}}, \apjs, 159, 60. \eprint{arXiv:astro-ph/0502378}

\bibitem[{{Suntzeff} et~al.(1988){Suntzeff}, {Heathcote}, {Weller}, {Caldwell},
  \& {Huchra}}]{Suntzeff88}
{Suntzeff}, N.~B., {Heathcote}, S., {Weller}, W.~G., {Caldwell}, N., \&
  {Huchra}, J.~P. 1988, \nat, 334, 135

\bibitem[{{van den Bergh}(1965{\natexlab{a}})}]{vandenBergh65b}
{van den Bergh}, S. 1965{\natexlab{a}}, \aj, 70, 667

\bibitem[{{van den Bergh}(1965{\natexlab{b}})}]{vandenBergh65a}
--- 1965{\natexlab{b}}, \pasp, 77, 269

\bibitem[{{van den Bergh}(1966)}]{vandenBergh66}
--- 1966, \pasp, 78, 74

\bibitem[{{Wang} et~al.(2002){Wang}, {Wheeler}, {H{\"o}flich}, {Khokhlov},
  {Baade}, {Branch}, {Challis}, {Filippenko}, {Fransson}, {Garnavich},
  {Kirshner}, {Lundqvist}, {McCray}, {Panagia}, {Pun}, {Phillips}, {Sonneborn},
  \& {Suntzeff}}]{Wang02}
{Wang}, L., {Wheeler}, J.~C., {H{\"o}flich}, P., {Khokhlov}, A., {Baade}, D.,
  {Branch}, D., {Challis}, P., {Filippenko}, A.~V., {Fransson}, C.,
  {Garnavich}, P., {Kirshner}, R.~P., {Lundqvist}, P., {McCray}, R., {Panagia},
  N., {Pun}, C.~S.~J., {Phillips}, M.~M., {Sonneborn}, G., \& {Suntzeff}, N.~B.
  2002, \apj, 579, 671. \eprint{arXiv:astro-ph/0205337}

\bibitem[{{Westerlund}(1961)}]{Westerlund61}
{Westerlund}, B. 1961, \pasp, 73, 72

\bibitem[{{Xu} et~al.(1994){Xu}, {Crotts}, \& {Kunkel}}]{Xu94}
{Xu}, J., {Crotts}, A.~P.~S., \& {Kunkel}, W.~E. 1994, \apj, 435, 274

\bibitem[{{Xu} et~al.(1995){Xu}, {Crotts}, \& {Kunkel}}]{Xu95}
--- 1995, \apj, 451, 806

\bibitem[{{Zwicky}(1940)}]{Zwicky40}
{Zwicky}, F. 1940, Reviews of Modern Physics, 12, 66

\end{thebibliography}

\end{document}